\documentclass[sigconf]{acmart}

\AtBeginDocument{%
  \providecommand\BibTeX{{%
    \normalfont B\kern-0.5em{\scshape i\kern-0.25em b}\kern-0.8em\TeX}}}

\copyrightyear{2022}
\acmYear{2022}
\setcopyright{acmlicensed}
\acmConference[WWW '22] {Proceedings of the ACM Web Conference 2022}{April 25--29, 2022}{Virtual Event, Lyon, France.}
\acmBooktitle{Proceedings of the ACM Web Conference 2022 (WWW '22), April 25--29, 2022, Virtual Event, Lyon, France}
\acmPrice{15.00}
\acmISBN{978-1-4503-9096-5/22/04}
\acmDOI{10.1145/3485447.3512087}

\newcommand{\model}{{DURation}}
\usepackage{listings}
\usepackage{xcolor}
\usepackage{float}
\usepackage{multirow}
\usepackage{hyperref}
\hypersetup{
    colorlinks=true,
    linkcolor=blue,
    filecolor=magenta,      
    urlcolor=cyan,
}

\colorlet{punct}{red!60!black}
\definecolor{background}{HTML}{EEEEEE}
\definecolor{delim}{RGB}{20,105,176}
\colorlet{numb}{magenta!60!black}

\lstdefinelanguage{json}{
    basicstyle=\normalfont\ttfamily,
    numbers=left,
    numberstyle=\scriptsize,
    stepnumber=1,
    numbersep=8pt,
    showstringspaces=false,
    breaklines=true,
    frame=lines,
    backgroundcolor=\color{background},
    literate=
     *{0}{{{\color{numb}0}}}{1}
      {1}{{{\color{numb}1}}}{1}
      {2}{{{\color{numb}2}}}{1}
      {3}{{{\color{numb}3}}}{1}
      {4}{{{\color{numb}4}}}{1}
      {5}{{{\color{numb}5}}}{1}
      {6}{{{\color{numb}6}}}{1}
      {7}{{{\color{numb}7}}}{1}
      {8}{{{\color{numb}8}}}{1}
      {9}{{{\color{numb}9}}}{1}
      {:}{{{\color{punct}{:}}}}{1}
      {,}{{{\color{punct}{,}}}}{1}
      {\{}{{{\color{delim}{\{}}}}{1}
      {\}}{{{\color{delim}{\}}}}}{1}
      {[}{{{\color{delim}{[}}}}{1}
      {]}{{{\color{delim}{]}}}}{1},
}

\makeatletter
\newenvironment{subtheorem}[1]{%
  \def\subtheoremcounter{#1}%
  \refstepcounter{#1}%
  \protected@edef\theparentnumber{\csname the#1\endcsname}%
  \setcounter{parentnumber}{\value{#1}}%
  \setcounter{#1}{0}%
  \expandafter\def\csname the#1\endcsname{\theparentnumber.\Alph{#1}}%
  \ignorespaces
}{%
  \setcounter{\subtheoremcounter}{\value{parentnumber}}%
  \ignorespacesafterend
}
\makeatother
\newcounter{parentnumber}

\newtheorem{thm}{Theorem}

\newtheorem{definition}{Definition}

\usepackage{tikz}
\usepackage{verbatim}
\usetikzlibrary{trees}

\AtBeginDocument{%
  \providecommand\BibTeX{{%
    \normalfont B\kern-0.5em{\scshape i\kern-0.25em b}\kern-0.8em\TeX}}}
    
\begin{document}

\title{The Name of the Title is Hope}

\title{Deep Unified Representation for Heterogeneous Recommendation}


\author{Chengqiang Lu}
\affiliation{%
  \institution{Alibaba Group}
  \country{China}
}
\email{lulu.lcq@alibaba-inc.com}

\author{Mingyang Yin}
\affiliation{%
  \institution{Alibaba Group}
  \country{China}
}
\email{hengyang.ymy@alibaba-inc.com}

\author{Shuheng Shen}
\affiliation{%
  \institution{Ant Financial Services Group}
  \country{China}
}
\email{shuheng.ssh@antgroup.com}

\author{Luo Ji}
\affiliation{%
  \institution{Alibaba Group}
  \country{China}
}
\email{jiluo.lj@alibaba-inc.com}

\author{Qi Liu}
\affiliation{%
  \institution{University of Science and Technology of China}
  \country{China}
}
\email{qiliuql@ustc.edu.cn}

\author{Hongxia Yang}
\affiliation{%
  \institution{Alibaba Group}
  \country{China}
}
\email{yang.yhx@alibaba-inc.com}

\fancyhead{}
\renewcommand{\shortauthors}{Trovato and Tobin, et al.}

\begin{abstract}
\pdfoutput=1
Recommendation system has been a widely studied task both in academia and industry.
Previous works mainly focus on  homogeneous recommendation and little progress has been made for heterogeneous recommender systems. However, heterogeneous recommendations, e.g., recommending different types of items including products, videos, celebrity shopping notes, among many others, are dominant nowadays.
State-of-the-art methods are incapable of leveraging attributes from different types of items and thus suffer from data sparsity problems. And it is indeed quite challenging to represent items with different feature spaces jointly.
To tackle this problem, we propose a kernel-based neural network, namely \underline{d}eep \underline{u}nified \underline{r}epresent\underline{ation} (or DURation) for heterogeneous recommendation, to jointly model unified representations of heterogeneous items while preserving their original feature space topology structures. Theoretically, we prove the representation ability of the proposed model. Besides, we conduct extensive experiments on the real-world datasets. Experimental results demonstrate that with the unified representation, our model achieves remarkable improvement (e.g., 4.1\% \textasciitilde  34.9\% lift by AUC score and 3.7\% lift by online CTR) over existing state-of-the-art models. 
\end{abstract}


\maketitle
\section{Introduction}
In the information explosion era, recommendation systems become more and more important to link users with their potential interesting items. Recently, along with the rising number and increasing diversity of available items, many online platforms provide different kinds of content to enrich users' experiences. For example,  e-commerce platforms recommend user-generated-contents (UGC) along with products, e.g., Amazon \cite{McAuley2013HiddenFA} and  Alibaba\cite{cen2019representation}. Social networking platforms  recommend potentially interesting books, movies, and music, e.g., Douban \cite{lian2017cccfnet} and Facebook \cite{shapira2013facebook}.  

Traditional homogeneous recommendation models that deal with products alone are inapplicable in these scenarios
\cite{cremonesi2010performance, gomez2015netflix}.
Therefore, heterogeneous recommendation \cite{Guerraoui2017HeterogeneousRW}, which remains a largely unexplored area, has attracted massive interest in both academia and industry. 
 
Most recommendation models extract users' and items' latent representations implicitly or explicitly and then calculate the relevance between them via inner-product or other neural network methods, e.g., MLP. Traditional models, such as Collaborative Filtering  \cite{sarwar2001item} and Matrix Factorization \cite{ koren2009matrix, xue2017deep}, exploit the interactions between users and items in-depth to learn representations. However, modeling the observed interactions is insufficient to make suitable recommendations as the informative attributes of users and items are ignored. In the past decade, deep learning has achieved great success in the areas of computer vision \cite{He_2016_CVPR}, natural language processing \cite{vaswani2017attention}, and natural science \cite{senior2020improved} with its powerful representation learning ability. Therefore, it becomes increasingly ubiquitous in modern recommendation systems to exploit arbitrary continuous and categorical attribute features with  deep neural networks \cite{covington2016deep}. These models usually compare favorably to the aforementioned traditional models because they could leverage items and users' attribute features. Besides, these models could remedy the problem of inadequate labeled interaction data.

Although previous works have made much progress, recommendation on heterogeneous items is still challenging because different kinds of attributes do not share the same feature space. 

Therefore, former models that were designed for homogeneous item recommendation could not be applied directly. A na\"ive but common way to solve it is to merely use the interaction features or the same part of attribute features, where much useful information has to be dropped ~\cite{huang2013learning}. Another solution is to train one sub-model for each kind  respectively while maintaining shared user embeddings \cite{lian2017cccfnet}. However, in this setting, one kind of item's recommendation could not benefit from other items' information. 
Cross-domain recommendation ~\cite{tang2012cross, gao2013cross,  cantador2015cross} is a similar problem but it deals with the samples from different distributions but with the \textit{same} feature space. Heterogeneous Information Networks (HIN)  \cite{shi2018heterogeneous} characterizes auxiliary data by constructing and mining a heterogeneous graph, e.g., a graph contains multiple types of entities such as movies, actors, and directors. Nevertheless, items to be recommended in HIN-based methods are still homogeneous, e.g. movies. Consequently, these models are not suitable for heterogeneous recommendation. 

To address this heterogeneous recommendation problem, we propose the \underline{d}eep \underline{u}nified \underline{r}epresent\underline{ation} (DURation) for the heterogeneous item recommendation to make full use of various item attribute features. We first introduce a simple and easy-to-extend model architecture. Compared with previous studies, we design the \textit{mapping  module} to transform  different kinds of input  features into a unified feature space. To learn the unified representation, we propose the \textit{alignment loss} to reduce their dissimilarities inspired by the domain generation~\cite{muandet2013domain}. Besides, we prove that the proposed unified representation could capture the nuanced difference of items after the alignment. We also modify the CORAL loss \cite{sun2016deep} to preserve the topology of original feature spaces. At the end, we design an approximated implementation with both theoretical and empirical proofs of its effectiveness for large-scale deployment. Results of relatively comprehensive experiments in the real-world datasets demonstrate the superiority of the proposed model.

To summarize, we make the following contributions: 
\begin{itemize}
\item We formally define the problem of heterogeneous item recommendation, a general framework for various cross-domain recommendation problems, and we discuss practical challenges of large-scale deployment.
\item We design unified representations of heterogeneous items to leverage different attribute spaces simultaneously and alleviate the problem of data sparsity. In both theory and practice, we prove that such alignment can keep topologies of their original spaces.
\item Extensive experiments on the real-world dataset demonstrate the effectiveness of our proposed model.
\end{itemize}  
The remainder of this paper is as follows. After introducing some related works in Section 2, we elaborate on our proposed model architecture and the learning method for the unified representations in Section 3 and 4 respectively. We then perform experimental evaluations in Section 5 and conclude the paper in Section 6.

\section{Related Work} 

\label{relate}
In this section, we briefly review recent progress in the cross-domain recommendation and transfer learning, especially their insufficiencies to be directly applied to the heterogeneous recommendation. 

\subsection{Cross-domain Recommendation} 

A considerable amount of literature has been published on the problem of cross-domain recommendation systems \cite{cantador2015cross, Lu2013SelectiveTL, Cremonesi2011CrossDomainRS}. In the different literature, researchers have considered distinct definitions of \textit{domain}. For the sake of clarity, we follow the definition of \cite{muandet2013domain}. Given input space $\mathcal{X}$ and output space $\mathcal{Y}$, the domain is a joint distribution $\mathbb{P}_{XY}$ on $\mathcal{X}\times \mathcal{Y}$. This definition covers the attribute-level and type-level domain defined in \cite{Cremonesi2011CrossDomainRS} but does not include the item-level domain.  

Cross-domain recommendation systems have achieved great success. For instance, multi-view neural networks ~\cite{lian2017cccfnet, li2020ddtcdr, He2018AGC}, that construct separated sub-networks for each kind of item (so-called view) and let them share the same user embedding layer, were proposed and widely used. However, the items of one kind could barely benefit from the information of the others in these models. Besides, probabilistic topic models are often used to learn topic distributions from different kinds of items then build relevance of these topic distributions \cite{tang2012cross,Tan2014CrossDR}. Conversely, topic models require establishing linkages across items by text data such as item descriptions and user-generated text and could not utilize raw attributes such as numerical or categorical features. In addition, neighbor-based models are also popular in building relations between different users.
One category of these models leverages users' similarity based on the users' social network \cite{Xu2016CrossdomainIR} that are not suitable for the platform lacking social information such as e-commerce platform. Furthermore, they do not perform well for the users with a weak social bond. The second category of neighbor-based models builds user similarity based on user latent features extracted from user attributes and history behaviors \cite{Wang2018CrossDomainRF}. One drawback of them is their poor scalability for large-scale online systems. Another drawback is all these models suffer from data sparsity. 

Along with the aforementioned definition of the domain, these models could extract latent features of items sampled from different distributions $\mathbb{P}_X$. Nonetheless, they require the original features of items in the same input space $\mathcal{X}$, which is not consistent with the setting of the heterogeneous item recommendation.  Therefore, most cross-domain models could not be applied directly to the problem. Actually, we demonstrate that cross-domain recommendation could be considered as a particular case of the heterogeneous item recommendation in Section \ref{problem}. 

\subsection{Transfer Learning}  

To learn the representation of items in different spaces, transfer learning is a conventional powerful tool in multiple areas including natural language processing \cite{ruder2019transfer}, compute vision \cite{weiss2016survey}, natural science \cite{lee2021transfer} and  recommendation system \cite{Man2017CrossDomainRA, chen2021deep}. EMCDR (Embedding and Mapping framework for Cross-Domain Recommendation) uses a multi-layer perceptron to capture the nonlinear mapping function to transfer knowledge from the source domain to the target domain \cite{Man2017CrossDomainRA}. The main disadvantage of EMCDR is that it only takes the interaction matrices of different items as input and cannot exploit the attribute features. Along this line, \citeauthor{li2020ddtcdr} \cite{li2020ddtcdr} propose a novel model based on the mechanism of dual learning that transfers information between two kinds of items so that this approach could make the use of different attribute features. It learns the latent orthogonal mapping function to preserve similarities of users' preferences. MMT-Net (Multi-Linear Module Transfer Network) leverages contextual invariances to develop shared modules across different items ~\cite{Krishnan2020TransferLV}. 

Domain adaptation (also known as covariate shift) is another form of transfer learning \cite{Pan2010ASO} that could be used in both supervised and unsupervised problems. Domain adaptation was proposed primarily to deal with the inconsistency between training and test distributions. \citeauthor{wang2011heterogeneous} \cite{wang2011heterogeneous} construct linear mapping functions to link different feature spaces using manifold alignment with the aid of label. However, the linear mapping function is not sufficient to learn representations exquisitely. In a broader setting, domain generalization deals with the instances sampled from multiple sources, especially unseen sources. DICA (Domain-Invariant Component Analysis) was proposed to learn the transformation by minimizing the dissimilarity across domains using a kernel-based optimization algorithm. These machine learning techniques are successful in many applications while the high computation makes it not suitable for a web-scale recommendation. Moreover, as we mentioned before, domain adaptation (domain generalization) could barely handle the distributions from different feature spaces in the heterogeneous item recommendation.

Despite its success in extracting representation of different kinds of items, the transfer learning-based model suffers from several major drawbacks. First of all, the representations learned by these models are not aligned among items and lack the preservation of topology of origin feature spaces which makes it obstructed to make full use of interaction data of different types of items. 
Besides, most transfer learning-based models need to specify one source and one target, which leads to poor extensibility to multiple sources.

\section{Framework}  

\begin{figure*}
    \centering
    \includegraphics[width=0.9\textwidth]{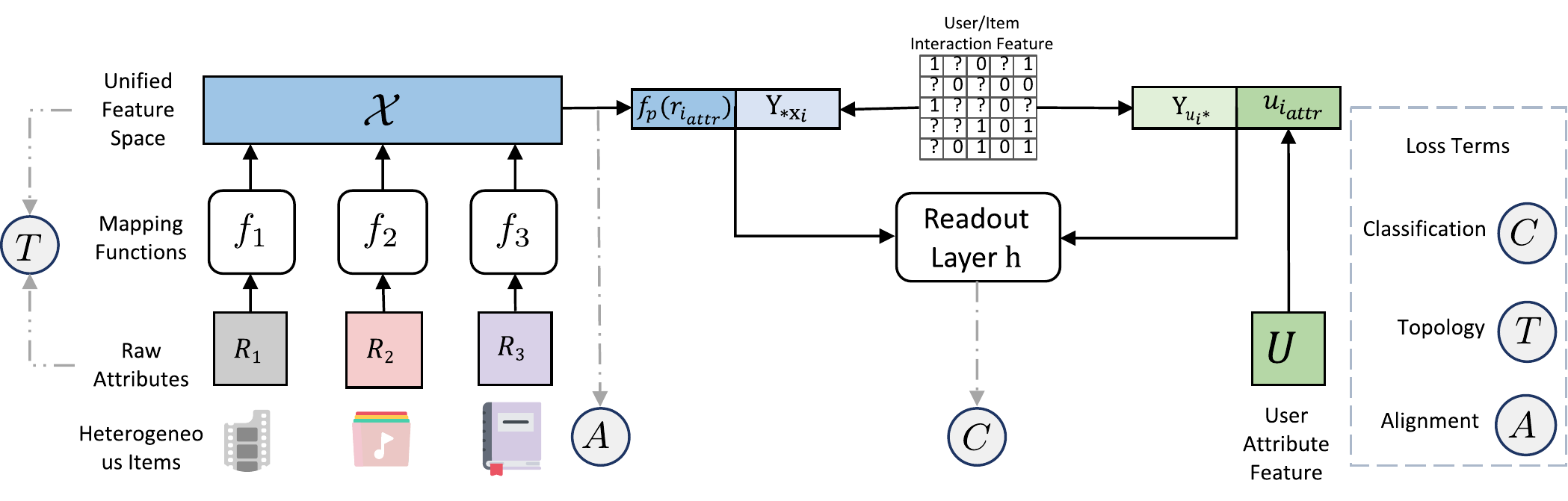}
    
    \caption{The framework of DURation. The left part is the learning of unified representation. For heterogeneous items, we construct multiple mapping functions to transform them into a unified feature space. The right part is the process of user and interaction features. After learning the user (green strip) and item (blue strip) embedding, we feed them to a readout layer to output the final prediction. The gray dash-dot lines indicate the computation of different loss terms.}
    \label{fig1}
    
\end{figure*}

In this section, we cover the specs of our proposed model. We formally introduce the heterogeneous recommendation problem followed by the detailed presentation of the proposed model.
\subsection{Problem Statement}   

\label{problem}
In the recommendation system, suppose we have a set of users $\mathbf{U}=\{u_1, u_2,\cdots, u_N\}$ and a set of items $\mathbf{R}=\{r_1, r_2,\cdots,r_M\}$, where $N$ and $M$ are the numbers of users and items respectively. We define a label matrix $Y \in \mathbb{R}^{N \times M}$ in which  $Y_{u,r}$ denotes the feedback of user $u$ on item $r$. The final task of a recommender is to predict the label for $u$ on item $i$ for which $Y_{u,r}$ is unobserved. 

In practice, we make our recommendation by learning latent representations of users $u$ and items $r$ from their interactions and attributes. In the setting of heterogeneous recommendation, the attributes of different kinds of items are in different feature spaces which makes it infeasible to directly apply the models that are designed for the homogeneous recommendation. Therefore, we need to construct mapping functions to transform the raw representations from their original feature spaces to a new space to obtain unified representation.   

Given the number of kinds of heterogeneous item $P$, let $\mathcal{R}_1 \cdots \mathcal{R}_P$ denote $P$ non-empty input spaces and $\mathcal{Y}$ an arbitrary output space. For each kind of item, let $\mathbb{P}_{R_{p}}$ denote a distribution on $\mathcal{R}_p$.  To map all $R_{p}$ into a new unified latent space  $\mathcal{X}$, we need build $P$ mapping functions $f_p: R_p\to X, p\in \{1\cdots,P\}$. Then, our goal is to produce an estimate function $g$ that convert the middle representation $X_p$ to the output $\hat{Y}$. Important mathematical notations used in the proposed model are listed in Table \ref{notation}.  

\begin{table}[ht] 
\centering
\scalebox{0.85}{
\begin{tabular}{llll}
\hline
Symbol        & Description      & Symbol        & Description           \\ \hline \hline
$u_{i_{attr}}$   & $i$th User's Features    & $\mathbf{U}$  & Set of Users          \\
$r^p_{i_{attr}}$   & $i$th Item's Features ($p$ type)    & $\mathbf{R}$  & Set of Items          \\
$Y$           & Label Matrix     & $\mathcal{Y}$ & Output Space          \\
$\mathcal{R}$ & Input Space      & $\mathcal{X}$ & Unified Feature Space \\
$f$           & Mapping Function & $\kappa$      & Kernel Function       \\
$\mathcal{H}$ & RKHS             & $\mu$         & Mean Map              \\ 
$\hat{Y}$  & Prediction Matrix  & $X_p$ & $p_{th}$ Item Representations \\\hline
\end{tabular}}
\caption{Mathematical Notations.} 
\label{notation}
\end{table}

Compared with existing works, the heterogeneous recommendation is a more general problem. If $\mathcal{R}_1=\mathcal{R}_2=\cdots=\mathcal{R}_P$, the problem degenerates to the cross-domain recommendation problem described in Section \ref{relate}. To pursue this further, if  $\mathbb{P}_{R_1}=\mathbb{P}_{R_2}=\cdots=\mathbb{P}_{R_p}$, this task degenerates to the common homogeneous recommendation.

\subsection{Model Architecture}  

\textbf{Overview.}
The overall framework of our model is a basic two-tower architecture that was first proposed for natural language tasks \cite{huang2013learning,neculoiu2016learning} and then becomes a popular approach in large-scale recommendation system \cite{yi2019sampling} because of its high efficiency. Considering a common setup for recommendation problems, given a set of users (queries) and items, we denoted the user feature vector by $\mathbf{u}$ and the item feature vector by $\mathbf{x}$. As illustrated in  Figure \ref{fig1}, the left and right towers encode user and item features respectively. Afterward, the item and user representations are fed into the readout layer to produce the final result.  

\textbf{User Representation.}  
For user $u_i$, to learn its representations $\mathbf{u}_i$, we use both the interactions between users and items and their attributes. Similar to DeepMF \cite{xue2017deep}, we represent interaction information by a high-dimensional vector of $Y_{u_i*}$ which denotes the $i_{th}$ user's rating across all items. Then, we concatenate item's interaction feature with attribute feature then pass them to a multilayer perceptron (MLP)  network $\mathbf{u_i} = \operatorname{MLP}(Y_{u_i*}\oplus u_{i_{attr}}).$ In our architecture, the MLP adopts ReLU as the activation function at the output layer. 

\textbf{Item Representation.} 
The calculation of item representation is analogous to user representation which also consists of interaction features and attribute features. However, in the heterogeneous recommendation, attribute features of items are from different distributions $\mathbb{P}_{\mathcal{R}_1},\cdots,\mathbb{P}_{\mathcal{R}_P}$ in different feature spaces $\mathcal{R}_1,\cdots,\mathcal{R}_P$ so that we need to transform them to the same space first. We aim to build $P$ mapping functions $f_p:\mathcal{R}_p\times \mathbb{R}^{d_p}\rightarrow\mathbb{R}^d$ where $d_p$ is the dimension of attribute feature space of the $p_{th}$ kind of items and $d$ is the dimension of the unified feature space. Here, we define the mapping function $f_p$ as an MLP too. After the transformation, we concatenate the transformed attribute feature vector with interaction feature vector then feed them into another MLP $\mathbf{x_i}  = \operatorname{MLP}(Y_{*x_i}\oplus f_p(r_{i_{attr}})).$

\section{Unified Representation Learning}     
The main impediment of applying existing models to the heterogeneous item recommendation is that the attributes features of different kinds of items are in different feature spaces. Simply dropping them leads to the loss of important side information of items, so that the learning of unified representations is the overarching task in the heterogeneous recommendation. There are three challenges here: firstly, the distribution $\mathbb{P}_{X_p}$ of unified representations of different kinds of items should be aligned; secondly, the new representations $X_p$ should preserve the topology of the original feature space; thirdly, we aim to keep the functional relationship $\mathbb{P}(Y|R_p)$ in the transformed feature space fixed so that the final deep features could be discriminative enough to train a strong recommendation system.     
\subsection{Alignment}    
 Intuitively, the performance of back-end model $h$ which works after learning unified representation depends in part on the dissimilarity of the distributions $\mathbb{P}_{X_1}, \mathbb{P}_{X_2}, \cdots, \mathbb{P}_{X_P}$. Besides, lack of feature distribution alignments can lead to overfitting when data is sparse \cite{mayer2019adversarial, wang2011heterogeneous}. Moreover, closely aligned feature distributions will make it tractable for downstream tasks such as item clustering. Therefore, alignment is one of the most essential parts in the learning of unified representations.   
 
To solve the problem of alignment, we propose to reduce the dissimilarity between feature spaces by minimizing the distance between empirical distributions of samples and these of their corresponding transformed ones by the map- ping functions.. Thus, the first step is to define the representation of distributions. Embedding probability distributions into reproducing kernel Hilbert spaces (RKHS) was introduced to resolve this issue \cite{berlinet2011reproducing}. This embedding represents any distribution as a mean element in an RKHS \cite{smola2007hilbert}. Given a kernel $\kappa$, we define the \textit{mean map} of the distribution $\mathbb{P}_{X_p}$ as 
\begin{equation}
    \mu[\mathbb{P}_{X_p}]:=\int_{\mathcal{X}} \kappa(x,\cdot)d\mathbb{P}_{X_p} (x).
\end{equation}  

Here $\mathbb{P}_{X_p}$ is the distribution of $p_{th}$ kind of items after transformation and $\mathcal{X}$ is the unified feature space. In our model we choose the Gaussian Radial Basis Function (RBF) kernel $\kappa(x, x')=\exp(-\lambda \|x-x'\|^2)$ due to its universality, which means for every function $f$ induced by $\kappa$ and every $\varepsilon > 0$ there exists a function $g$ induced by $k$ with $\|f-g\|_{\infty}\leq \varepsilon$. This property ensures that all continuous functions can be approximated by certain kernel expressions \cite{steinwart2001influence} and benefits the embedding of distributions which will be discussed later.  

By utilizing the mean map and RBF kernel, the distance between distributions is measured as the inner product$\langle \mu[\mathbb{P}_{X_p}], \mu[\mathbb{P}_{X_{p'}}]\rangle$.
To prove the suitability of this measure, let us introduce the \textit{characteristic kernel} first \cite{sriperumbudur2011universality}:

\begin{definition}[Characteristic kernel]
\label{ckernel} 
A bounded measurable kernel, $\kappa$ is said to be characteristic if $\mu \rightarrow \int_{\mathcal{X}}\kappa(x,\cdot)d\mu(x)$ is injective, where $\mu$ is a Borel probability measure on $\mathcal{X}$.
\end{definition}  

Let $\mu$ be the mean map using RBF as the kernel function. The following theorem shows that the inner product between RBF kernel mean embedding is suitable as a measure of the distance between distributions.

\begin{subtheorem}{thm}\label{thm:one}
\begin{thm}\label{thm:oneA} 
For any two distributions $\mathbb{P}$ and $\mathbb{Q}$ in $\mathcal{X}$, $\mu[\mathbb{P}] = \mu[\mathbb{Q}]$ if and only if  $\mkern9mu \mathbb{P} = \mathbb{Q}$.
\end{thm}
\begin{thm}\label{thm:oneB}
There exist two distinct distributions $\mathbb{P}$ and $\mathbb{Q}$ defined on $\mathcal{X}$ satisfying $\langle \mu[\mathbb{P}] , \mu[\mathbb{Q}]   \rangle_{\mathcal{H}} < \varepsilon$ for any arbitrarily small $\varepsilon > 0$.
\end{thm}
\end{subtheorem}

The Theorem \ref{thm:one} demonstrates that distributions are close in embedding space when their differences occur at higher frequencies and this embedding could capture the nuanced difference. We provide the proof of this theorem in Supplementary Material \ref{ap:proof1}. \ After embedding the distributions, a popular method to measure the dissimilarity between distributions is using the \textit{distributional variance} \cite{muandet2013domain}:   

\begin{definition}[Distributional variance]
Given a set of distributions $\mathcal{P} = \{\mathbb{P}_{X_1},\mathbb{P}_{X_2},\cdots,\mathbb{P}_{X_P}\}$, embed them to RKHS $\mathcal{H}$ and then define the $P\times P$ Gram matrix $G$ with entries $G_{ij}=\langle\mu_{\mathbb{P}_{X_i}}, \mu_{\mathbb{P}_{X_j}} \rangle_{\mathcal{H}}$. Let $\Sigma=G-\mathbf{1}_{P} G-$ $G \mathbf{1}_{P}+ \mathbf{1}_{P} G \mathbf{1}_{P}$. The distributional variance is     
\begin{equation}
    \mathbb{V}_{\mathcal{H}}(\mathcal{P})=\frac{1}{P} \operatorname{tr}(\Sigma)=\frac{1}{P} \operatorname{tr}(G)-\frac{1}{P^{2}} \sum_{i=1}^{P}\sum_{j=1}^{P} G_{i j}.
\end{equation}
\end{definition}
Here $\mathbf{1_{P}}$ denotes a $P \times P$ all-ones matrix. Given $P$  sample sets $\mathcal{S}=\{S^i\}_{i=1}^P$ drawn from $\mathbb{P}_{X_1},\mathbb{P}_{X_2},\cdots,\mathbb{P}_{X_P}$,  previous study \cite{muandet2013domain} proved that the empirical estimator $\widehat{\mathbb{V}}_{\mathcal{H}}(\mathcal{S})=\frac{1}{P} \operatorname{tr}(\widehat{\Sigma})$ is a consistent estimation of $\mathbb{V}_{\mathcal{H}}(\mathcal{P})$ where $\widehat{\Sigma}$ is obtained from the Gram matrix 
\begin{equation}
     \widehat{G}_{i j}:=\frac{1}{N_{i} \cdot N_{j}} \sum_{k=1}^{N_{i}} \sum_{l=1}^{N_{j}} \kappa\left(x_{k}^{i}, x_{l}^{j}\right).
\end{equation} 
Here $i, j$ is the index of distributions, and $k, l$ is the index of the samples in their distribution. $N_i$ denotes the number of samples drawn from $i_{th}$ distributions.
To investigate this further, aiming to minimize $\widehat{\mathbb{V}}_{\mathcal{H}}(\mathcal{S})$, we define the \textit{alignment loss}  
\begin{equation}
\label{eq:align}
    A = \sum_{i=1}^{P}\sum_{j=1}^{P}\sum_{k=1}^{N_i} \sum_{l=1}^{N_j} A_{ijkl},\quad     
    A_{i j k l}=\left\{\begin{array}{cc}     
\frac{P-1}{P^{2} N_{i}^{2}} \kappa \left(x_{k}^{i}, x_{l}^{j}\right) & i=j \\
\frac{-1}{P^{2} N_{i} N_{j}} \kappa \left(x_{k}^{i}, x_{l}^{j}\right) & i \neq j 
\end{array}\right.
\end{equation}

The derivation process from $\widehat{\mathbb{V}}_{\mathcal{H}}(\mathcal{S})$ to Equation \ref{eq:align} is investigated in Supplementary Material \ref{ap:proof}. Because the complexity of Equation \ref{eq:align} is high, we make an approximation which is discussed in Section \ref{obj}.

\subsection{Topology Preservation}   
During the construction of the mapping functions, we aim to preserve the topology of the original space for the raw feature of each kind of item. Given two items that belong to the same category $i_{th}$ , their raw and mapped embeddings are $\mathbf{r}^i_k, \mathbf{r}^i_l$ and $\mathbf{x}^i_k, \mathbf{x}^i_l$ respectively. Basically, if $\mathbf{r}^i_k$ and $\mathbf{r}^i_l$ are similar in their original spaces, the $\mathbf{x}^i_k$ and $\mathbf{x}^i_l$ should also close in the transformed feature space and vice versa. To solve the problem, several attempts have been made to minimize the distances between pairs of items in the new space with their similarity in raw space as an adjusted rate \cite{wang2011heterogeneous}. However, this approach suffers from the choice of similarity metric: Gaussian kernel similarity could not ensure the dissimilar items are well-separated from each other in the new space; Euclidean distance varies greatly which makes it intractable in practice; adjusted cosine similarity performs unstable in experiments.

As a remedy, we propose a modification of the CORAL loss \cite {sun2016deep} which is defined as the distance between the second-order statistics. Suppose we are given $i_{th}$ kind training examples $D_{R^i}=\{\mathbf{r}^i_k\}, \mathbf{r}\in\mathbb{R}^d$ where $d$ is embedding dimension. $D_{X^i}=\{\mathbf{x}^i_k\}, \mathbf{x}\in\mathbb{R}^{d'}$  denotes the samples after the transformation. Therefore, the covariance matrices of the raw and transformed data are given by :   
\begin{equation}
\begin{aligned}
C_{R^{i}} &=\frac{1}{d-1}\left(D_{R^{i}} D_{R^{i}}^{\top}-\frac{1}{d}\left(\mathbf{1}^{\top} D_{R^{i}}^{\top}\right)^{\top}\left(\mathbf{1}^{\top} D_{R^{i}}^{\top}\right)\right) \\
C_{X^{i}} &=\frac{1}{d^{\prime}-1}\left(D_{X^{i}} D_{X^{i}}^{\top}-\frac{1}{d^{\prime}}\left(\mathbf{1}^{\top} D_{X^{i}}^{\top}\right)^{\top}\left(\mathbf{1}^{\top} D_{X^{i}}^{\top}\right)\right),
\end{aligned}
\end{equation}
where $\mathbf{1}$ is a column vector with all elements equal to 1. Then we define the \textit{topology loss} as: 
\begin{equation}\label{eq:topo}
        T = \sum_{i=1}^P\frac{1}{4N_i}\|C_{R^i} - C_{X_i}\|^2_F,
\end{equation}  
where $\|\cdot\|^2_F$ denotes the squared matrix Frobenius norm. 
\subsection{Objective Function}   
\label{obj}
Because we model the recommendation as a classification problem, we utilize the Binary Cross Entropy Loss (BCE Loss) as the \textit{classification loss}:  
\begin{equation}
    \label{eq:cls}
    C = -\frac{1}{M}\sum_{i=1}^M  y_i \log(\hat{y}_i) + (1 - y_i)  \log(1-\hat{y_i}),
\end{equation}  
where $M$ is the number of all labeled interactions between users and items and $y_i$ is the user's feedback to the item in one interaction. $\hat{y}_i=g(x_i)$ indicates the predictive value where $h$ is the readout function mentioned in Section 3.2.  

We desire our model to simultaneously achieve the aforementioned three goals in the new space and obtain a unified representation. Consequently, combining   Equation (\ref{eq:align}), (\ref{eq:topo}) and (\ref{eq:cls}), the overall objective function to be minimized is
\begin{equation}
\label{opt}
\setlength{\abovedisplayskip}{1pt}
\setlength{\belowdisplayskip}{3pt}
    L= C + \alpha A + \beta T,
\end{equation} 
where $\alpha$ and $\beta$ are hyperparameters. We optimize the Equation \ref{opt} with Adam optimizer.   
For each distribution, we randomly sample $B$ samples to calculate the mini-batch stochastic gradient   
$\nabla L_{PB} = \frac{1}{B} \sum_{i=1}^P (\nabla C_B + \alpha \nabla A_B + \beta \nabla T_B)$.   
As sample indexes to estimate the loss terms in Equation \ref{eq:align}, \ref{eq:topo} and \ref{eq:cls} are all independent, $\nabla L_{PB}$ is an unbiased estimation of $\nabla L$, which guarantees a $\frac{1}{PB\epsilon^2}$ convergence rate. Therefore, this approximation ensures that our model could be applied to the web-scale recommendation.

\textbf{Summary.}  
With  latent representations of users and items, a readout function $h$ is applied for the final prediction, which is denoted as $\hat{y}_{ij}= h( \mathbf{u}_i,\mathbf{x}_j)$. The choice of $h$ can be diverse, such as bidirectional encoder representations \cite{sun2019bert4rec,kang2018self} which is capable of handling sequential recommendation, or deep ranking models \cite{li2019multi} to better capture high order interaction. In the consideration of online efficiency for large-scale system, we use inner product of two vectors as $h( \mathbf{u}_i,\mathbf{x}_j) = \langle \mathbf{u}_i,\mathbf{x}_j \rangle$.

\section{Experiments}
In this section, we conduct extensive experiments on a real-world public dataset and on a real-world large-scale online recommendation system. Besides, the code of our proposed model and baselines has been published to GitHub \footnote{https://github.com/geekinglcq/HRec} so that everyone could validate the experimental results using a one-line code with proper configuration.  For more details about reproducibility, such as the dataset, code usage, set-up, and hyperparameters, please refer to the Supplementary Material \ref{ap:rep}.

\subsection{Dataset}
In previous work, synthetic datasets (e.g. sample data from the homogeneous database such as Google Books, IMDB, and Spotify respectively and then mixed them up) are widely used \cite{Man2017CrossDomainRA, li2020ddtcdr}. However, these datasets are simply combinations of some well-known homogeneous datasets and do not correspond with reality. Therefore, we evaluated our model on the real-world Douban dataset. Douban\footnote{www.douban.com} is a popular Chinese social networking platform that allows users to create content related to movies, books, and music. Here we convert the recommendation problem in the Douban dataset to a binary classification problem (like or not like). The detailed statistics of this dataset are shown in Table \ref{tb:dataset}. Besides, we also deploy our model into production and conduct online A/B test experiments which will be explained in Section \ref{abtest}.

\begin{table}[htbp]
\caption{Statistics of the Douban dataset.}  
\centering
\label{tb:dataset}
\begin{tabular}{lllll}
\hline
      & \# Item & \# Rate & \# User & \# Attribute \\ \hline \hline
Book  & 150955  & 356001  & 25496   & 11           \\
Music & 102571  & 269760  & 25426   & 8            \\
Movie & 50299   & 470479  & 31664   & 10           \\
All   & 303825  & 1096240 & 36585   & /            \\ \hline
\end{tabular} 

\end{table}

\subsection{Baselines}
\label{baseline}
\subsubsection{General Models}        

The first group consists of three general models. They merely use the interaction information between user and items and drop the attributes of items. 
\begin{itemize}
    \item \textbf{DeepMF} \cite{xue2017deep} is a neural network enhanced matrix factorization model.
    \item \textbf{FISM} \cite{kabbur2013fism} is an item-based model that learns the item-item similarity for the recommendation.  
    \item \textbf{NAIS} \cite{he2018nais} is a deep attention model that is capable of distinguishing which historical items in a user profile are more important.
\end{itemize}

\subsubsection{Contextual Models}   

The second group contains six contextual models that could exploit the attributes of homogeneous items. To apply them in the heterogeneous recommendation, we feed them with the overlapping features of different kinds of items such as the average rate of items. Besides, we also run these models in each kind of homogeneous item for comparison.
\begin{itemize}
    \item \textbf{DeepFM} \cite{guo2017deepfm} combines the power of factorization machines for recommendation and deep learning for feature learning. 
    \item \textbf{xDeepFM} \cite{lian2018xdeepfm} uses the compressed interaction network to generate combinatorial features and merges them into a classic deep neural network.
    \item \textbf{AFM} \cite{xiao2017attentional} is the attentional factorization machine that could learn the importance of different feature interactions. 
    \item \textbf{DSSM} \cite{huang2013learning} is one of the first proposed two-tower models used in the recommendation system.
    \item \textbf{Wide\&Deep} \cite{cheng2016wide} is an approach that uses wide linear models and deep neural networks to combine the benefits of memorization and generalization. 
    \item \textbf{AutoInt} \cite{song2019autoint} is a deep model that could learn the high-order feature interactions of input features via self-attentive neural networks. 
\end{itemize}  

\subsubsection{Heterogeneous Models}    

The third group contains two cross-domain models that could handle heterogeneous items mentioned in related work.    
\begin{itemize}
    \item \textbf{CCCFNet} \cite{tang2012cross} combines the collaborative filtering feature and content-based feature in a multi-view neural network. 
    \item \textbf{DDTCDR} \cite{li2020ddtcdr} develops a novel latent orthogonal mapping to extract user preferences over multiple sources while preserving the relations between users across different latent spaces. We extend the origin DDTCDR model to a multiple-source version so that it could be applied to the heterogeneous item recommendation.
\end{itemize}

\subsection{Effectiveness}    
\label{eff}
\subsubsection{Experiments on Public Dataset}
\leavevmode   
We conduct effectiveness experiments in all of the baselines and our proposed model. For the contextual models, we run them in both heterogeneous and homogeneous settings. In the former setting (marked with He), these models use the overlapping attributes of different kinds of items as the contextual features; in the latter setting (marked with Ho), these models use full attributes of homogeneous items.   

Table \ref{tb:ef} shows the recommendation performance of baselines and our model. We have the following main observations. 

    Our model achieves the highest AUC scores in all kinds of items. In the heterogeneous setting, our model could exploit more attribute features of items. Besides, our model compares favorably to homogeneous contextual models due to its benefit from the data of all kinds of items. 
    Basically, contextual models perform better than general models, and contextual models in the homogeneous setting perform better than them in the heterogeneous setting. Besides, the heterogeneous models (CCCFNet and DDTCDR) outperform the contextual models. To pursue this further, our model achieves significant improvements over other models. We could clearly see the positive effect of utilization of attribute features: general models drop all attributes of items; heterogeneous contextual models merely use the common attributes of items; homogeneous models could manipulate all attribute of one kind items; our model enables the usage of all attributes of different kinds of items. 
    As the interaction matrix of items gets more sparse, the superiority of our model becomes more salient. Take the book item as an example, its density of the interaction matrix (i.e., $\text{\# Rate}/(\text{\# Item}\times \text{\# User})$) is much lower than other two kinds of items (see Table \ref{tb:dataset} ). Table \ref{tb:ef} illustrates that, when recommending books, the advantage of \model~ is more significant, which demonstrates that our model profits from interaction data of the heterogeneous items.
 \begin{figure}
    \centering
    \includegraphics[width=0.51\textwidth,trim={90 0 0 0}]{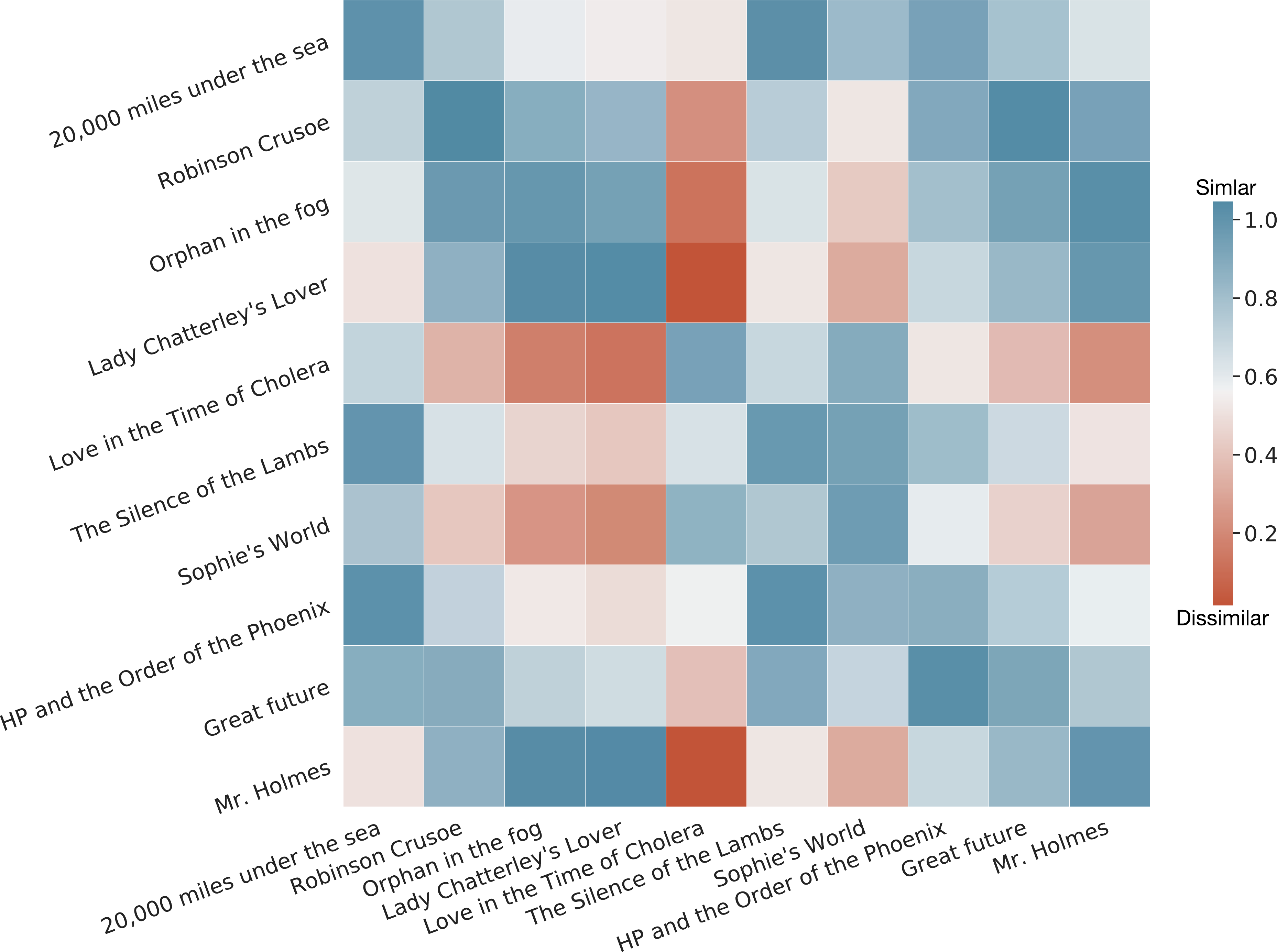}
    \caption{Book-Movie Similarity Heatmap. The horizontal axis represents ten books while the vertical axis represents their movie adaptations. The color of the block indicates the similarity between the corresponding book and movie. The bluer the higher the similarity and vice versa.}
    \label{fig:heatmap}
     
\end{figure}

\begin{table*}[]
\caption{Comparative results with baselines and \model. Different categories of models are separated by vertical lines. He indicates the heterogeneous setting and Ho indicates the homogeneous setting. The highest scores are marked in bold.} 

\label{tb:ef}
\scalebox{0.85}{
\begin{tabular}{l|lll|llllllllllll|ll|l}
\hline
AUC   & NAIS   & DeepMF & FISM  & \multicolumn{2}{l}{DeepFM} & \multicolumn{2}{l}{xDeepFM} & \multicolumn{2}{l}{AFM} & \multicolumn{2}{l}{DSSM} & \multicolumn{2}{l}{Wide\&Deep} & \multicolumn{2}{l|}{AutoInt} & CCCFNet & DDTCDR & \model~   \\ \hline \hline
      & He     & He     & He    & He           & Ho          & He           & Ho           & He         & Ho         & He          & Ho         & He             & Ho            & He            & Ho           & He      & He     & He    \\
Book  & 0.554  & 0.588  & 0.571 & 0.721        & 0.737       & 0.721        & 0.732        & 0.686      & 0.73       & 0.519       & 0.545      & 0.722          & 0.739         & 0.722         & 0.738        & 0.741   & 0.734  & \textbf{0.766} \\
Music & 0.563  & 0.631  & 0.576 & 0.756        & 0.782       & 0.759        & 0.769        & 0.709      & 0.769      & 0.543       & 0.571      & 0.759          & 0.784         & 0.76          & 0.784        & 0.786   & 0.776  & \textbf{0.818} \\
Movie & 0.651  & 0.656  & 0.656 & 0.76         & 0.777       & 0.761        & 0.78         & 0.739      & 0.771      & 0.582        & 0.707      & 0.763          & 0.78          & 0.763         & 0.78         & 0.782   & 0.774  & \textbf{0.792} \\
All   & 0.592 & 0.647  & 0.596 & 0.77         &      /       & 0.77        & /            & 0.708      & /          & 0.569       & /          & 0.771          & /             & 0.771         & /            & 0.787   & 0.78   & \textbf{0.804} \\ \hline
\end{tabular}}

\end{table*}

Next, we conduct experiments to demonstrate that unified representations could benefit from attributes of all kinds of items. The representation of one kind of item also benefits from the information of other items. Aiming to show the connection between representations of different kinds of items, we conduct a case study. Firstly, we select some famous books and their movie adaptation. Then we obtain their unified representations and calculate their pair-wise similarity to construct the correlation matrix. Figure \ref{fig:heatmap} is the heatmap of the correlation matrix. The vertical axis represents the movies and the horizontal axis indicates their original books. Each block shows the similarity between the book and the movie. The bluer the block is, the more similar the book and the movie are. When the color becomes redder, the book and the movie are more distinct. We can see that the block in the diagonal is bluer than the other parts of the figure dramatically. It shows that the unified representations could capture the adaptation relationship of books and movies. Besides, it is reasonable to see that <Love in the Time of Cholera> is more distinct from other books/movies because this book and its film adaptation are relatively unpopular. As a summary, we could draw the conclusion that the unified representations of different kinds of items are integrated well in the new feature space.

\begin{table}[h]
\centering
\scalebox{0.95}{
\begin{tabular}{llllll}

\hline
            & PV gain  & UV gain & CTR gain \\ \hline \hline 
All         & +4.26\%  & +1.75\% & +3.67\%  \\
Merchandise & -1.05\%  & +1.21\% & +1.25\%  \\
Article     & +10.67\% & +4.05\% & +2.62\%  \\
QA          & +6.78\%  & +1.01\% & +3.04\%  \\
Video       & +1.5\%   & +1.31\% & +5.70\%   \\ \hline
\end{tabular}}
\caption{Online experimental results.}
\label{online_2}
\end{table}

\subsubsection{Online A/B Testing.} 
\leavevmode \\  

\label{abtest}
In addition to the offline experiments in the public dataset, We also apply ~\model\ model in a real-world large-scale online recommendation system. In this scenario, there are four types of items. They are merchandise, videos, QAs(Question-Answer), and articles. In this recommendation system, the Daily Active Users (DAU) are about six million. We train our model with the data set that sampled from user behaviors in one day and deploy this model in the next day's recommendation. The amount of data in our training set here is about 50 million.   

The original algorithm in this scenario is a CCCFNet-based \cite{tang2012cross} model. Except for the basic components of the  naive CCCFNet, this model also utilizes self-attention to extract the user representation from the sequence of user behavior.  

In the offline experiment, the AUC of our model is 0.675 while the AUC of the baseline is 0.667. Compared with this baseline, our proposed method contributes to 3.67\% CTR gain, 4.26\% PV gain, and 1.75\% UV gain in the online A/B test. Especially, our model performs  relatively better in the setting of cold-start, which leads to a 27.22\% CTR increase in the recommendation of new items and a 2.59\% CTR increasineor new users. The definition of a new user/item is the user/item that had no activity in the past 30 days. Table \ref{online_2} illustrates the detailed results of online experiments including the performance of our model in different types of items.

\subsection{Ablation Study}

\begin{figure}
    \centering
    \includegraphics[width=0.37\textwidth]{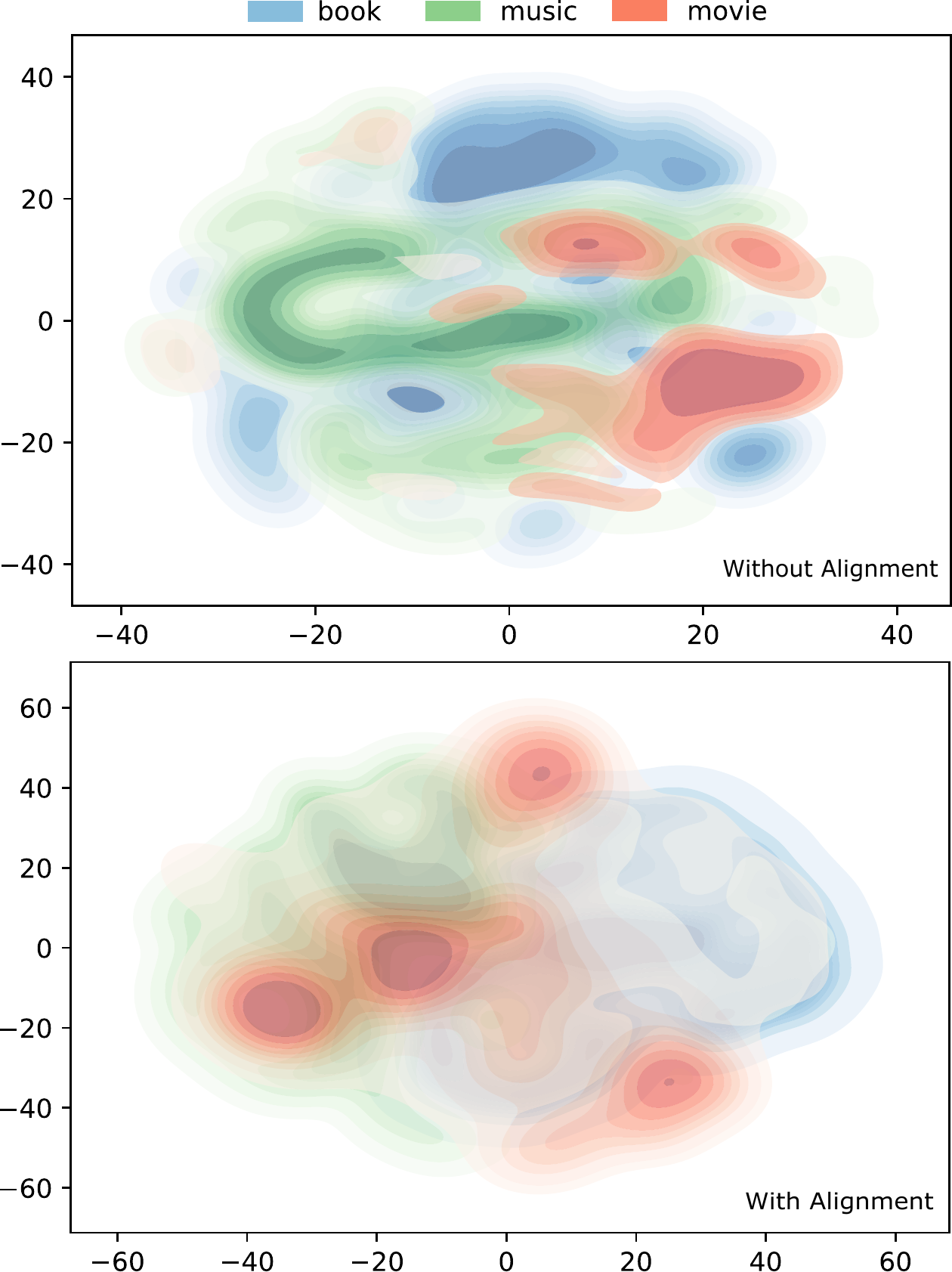} 
    
    \caption{Visualization of item representations using t-SNE. Different colors indicate different kinds of items. The upper and under sub-figures demonstrate the distribution of representations learned in models with and without alignment loss respectively. It shows that the alignment improves the smoothness of integration among distinct distributions.}
    \label{fig:alignemnt}
     
\end{figure}
\begin{figure}
    \centering
    \includegraphics[width=0.37\textwidth]{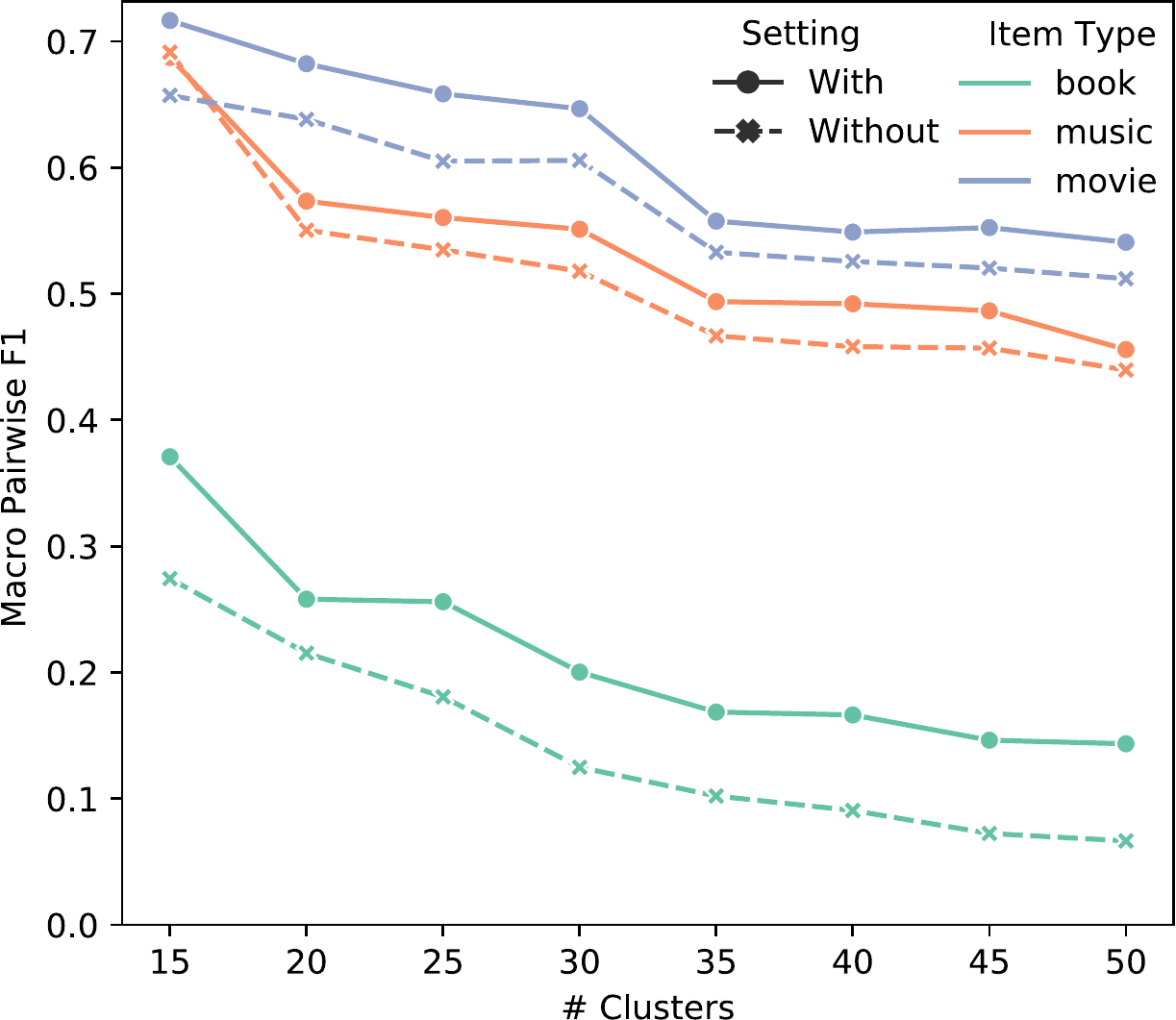}
    
    \caption{Clustering results with a different number of clusters. The dotted line indicates the result of the model with topology loss, and the dashed line with a cross mark indicates the model without topology loss. A higher macro pairwise F1 score reveals the increased ability to preserve the original space's topology.}
    \label{fig:cluster}

\end{figure}

In this subsection, we conduct experiments to demonstrate how each part of our model affects the learning of the unified representation. To validate the necessity of the alignment loss and topology loss, we remove them and then test our model. The results show that the AUC score drops from 0.804 to 0.798 in the full dataset. As the cold-start problem is one of the most significant challenges in the recommendation system, we also test the performance of our model in the cold-start setting. In this setting, the user in the test set is inactive and we only reserve up to three samples of these users each in the training set. Removing $A$ and $T$ decreases the AUC score remarkably (from 0.675 to 0.649)  in the cold-start setting. For detailed results, please refer to Table \ref{ablation}.  

To get a better understanding of alignment loss $A$, we use t-SNE \cite{van2008visualizing, chan2019gpu} to do dimensionality reduction and then plot the distribution of our unified representations using kernel density estimate (KDE). Figure \ref{fig:alignemnt} illustrates the results of visualization. The upper plot is the representations learned by the model without alignment loss and the underplot is the output of our model with alignment loss. Different color indicates a different kind of items. We can clearly see that the cliques in the upper sub-figure are more distinct and the distributions in the under sub-figure are well-mixed. It proves that the alignment loss helps minimize of dissimilarity of different kinds of items in the unified feature space so that the unified representation becomes more tractable.  

\begin{table}[h]
\centering
\begin{tabular}{llllll}
\hline
Setting                     & Model              & All   & Book  & Music & Movie \\ \hline\hline 
\multirow{2}{*}{Cold-start} & Duration           & 0.576 & 0.707 & 0.667 & 0.675 \\
                            & Remove $A$ and $T$ & 0.565 & 0.69  & 0.619 & 0.649 \\ \hline
\multirow{2}{*}{Full data}  & Duration           & 0.804 & 0.766 & 0.819 & 0.792 \\
                            & Remove $A$ and $T$ & 0.798 & 0.759 & 0.817 & 0.79  \\ \hline
\end{tabular}
\caption{Performance result in ablation study.}
\label{ablation}

\end{table}

We also design an experiment to validate the competency of the topology loss $T$. Firstly, for each item kind, we cluster the raw item representations using the K-Means algorithm, then assign each item a pseudo label. Afterward, we cluster the items with their unified representations learned in two settings (with/without topology loss). Then we calculate the macro pairwise F1 score of the two cluster results generated by raw representations and unified representations.  For each pair of items, if their raw representations belong to the same cluster, this pair is regarded as a positive sample. In that situation, if their deep unified representations also belong to the same cluster, it leads to a true-positive prediction. Analogously, we could count the number of true negatives, false positives, and false negatives. Accordingly, the pairwise precision and pairwise recall are easy to be induced by the definition in the simple binary classification context. Figure \ref{fig:cluster} presents the pairwise F1 score of models in different settings. We can clearly see that if removed the topology loss $T$, the F1 score goes down. This indicates that the topology loss indeed boosts the ability  to preserve the topology information of origin feature space.

\section{Conclusion}  

In this work, we propose a novel deep unified representation for the heterogeneous recommendation. We leverage the strengths of attributes of different kinds of items to learn the unified exquisite representations. We design an easy-to-extend framework and introduce the alignment loss, topology loss, and classification loss in the learning of unified representation. In theory and in practice, we prove that the representation could align different transformed distributions while preserving the topology of raw feature spaces. Furthermore, our model achieves much better performance compared with baselines on real-world offline and online datasets.

\bibliographystyle{ACM-Reference-Format}
\bibliography{main}

\clearpage
\appendix
\section{Supplementary Material}

\subsection{Proofs}
\subsubsection{Proof of Theorem \ref{thm:one}}     

\label{ap:proof1}
\begin{proof}
We prove the Theorem \ref{thm:oneA} first. It is clear that $\mu[\mathbb{P}] = \mu[\mathbb{Q}]$ if  $\ \mathbb{P} = \mathbb{Q}$. If  $ \mathbb{P} \neq \mathbb{Q}$, then  $\operatorname{MMD} (C(\mathcal{X}),\mathbb{P}, \mathbb{Q}) > 0$ based on the Lemma 1 in \cite{gretton2008kernel} where $\operatorname{MMD}$ is maximum mean discrepancy and $C(\mathcal{X})$ is the space of bounded continuous functions on $\mathcal{X}$. Therefore, we could prove the converse by showing that  $\operatorname{MMD} (C(\mathcal{X}),\mathbb{P}, \mathbb{Q}) =D$  for some $D>0$ implies  $\mu[\mathbb{P}] \neq \mu[\mathbb{Q}]$.  

Because the RBF kernel $k$ is universal, its associated RKHS $\mathcal{H}$ is also universal. Let $\mathcal{F}$ be the unit ball in $\mathcal{H}$. If $\operatorname{MMD}[C(\mathcal{X}), \mathbb{P}, \mathbb{Q}]=D,$ then there exists some $\tilde{f} \in C(\mathcal{X})$ for which $\mathbf{E}_{\mathbb{P}}[\tilde{f}]-\mathbf{E}_{\mathbb{Q}}[\tilde{f}] \geq D / 2 .$ Therefore, for $\epsilon=D / 8$, there exists $f*\in \mathcal{H}$ so that $\left\|f^{*}-\tilde{f}\right\|_{\infty}<\epsilon $ since $\mathcal{H}$ is dense in $C(\mathcal{X})$ with respect to the $L_\infty$ norm. Then, we have $\left|\mathbf{E}_{\mathbb{P}}\left[f^{*}\right]-\mathbf{E}_{\mathbb{P}}[\tilde{f}]\right|<\epsilon$ and consequently

\begin{equation*} \label{proof}
\begin{split}
 \left|\mathbf{E}_{\mathbb{P}}\left[f^{*}\right]-\mathbf{E}_{\mathbb{Q}}\left[f^{*}\right]\right| & > \left|\mathbf{E}_{\mathbb{P}}[\tilde{f}]-\mathbf{E}_{\mathbb{Q}}[\tilde{f}]\right|-2 \epsilon \\
     &>\frac{D}{4} >0.
\end{split}
\end{equation*}  

Then we obtain 
\begin{equation*}
\frac{\mathbf{E}_{\mathbb{P}}\left[f^{*}\right]-\mathbf{E}_{\mathbb{Q}}\left[f^{*}\right]}{ \left\|f^{*}\right\|_{\mathcal{H}}}
 \geq \frac{D}{4\left\|f^{*}\right\|_{\mathcal{H}}} >0.  
\end{equation*}  
Therefore, the definition of MMD establishes that $\operatorname{MMD}(\mathcal{F}, \mathbb{P}, \mathbb{Q}) > 0$. Since $\operatorname{MMD}^2(\mathcal{F}, \mathbb{P}, \mathbb{Q})=\|\mu[\mathbb{P}] - \mu[\mathbb{Q}]\|^2_\mathcal{H}$, we have $\mu[\mathbb{P}] \neq\mu[\mathbb{Q}]$. Finally, we can conclude that now  $\mu[\mathbb{P}]=\mu[\mathbb{Q}] \iff \mathbb{P}=\mathbb{Q}$. 

Since the Theorem \ref{thm:oneA} is proved, namely, the mean map $\mu \rightarrow \int_{\mathcal{X}}\kappa(x,\cdot)d\mu(x)$ is injective and hence the RBF kernel $k$ is characteristic by the Definition \ref{ckernel}. Along this line, we can construct $\mathbb{P} \neq \mathbb{Q}$ such that $\langle \mu[\mathbb{P}] , \mu[\mathbb{Q}]   \rangle_{\mathcal{H}} < \varepsilon$ for any arbitrarily small $\varepsilon > 0$ which is proved in previous work (\cite{sriperumbudur2010hilbert}, Theorem 9). 
\end{proof}

\subsubsection{Proof of Equation \ref{eq:align}}     

\label{ap:proof}
 \begin{proof}
From $\widehat{\mathbb{V}}_{\mathcal{H}}(\mathcal{S})$ to Equation \ref{eq:align}, we have  
\begin{equation}
    \begin{split}
    \widehat{\mathbb{V}}_{\mathcal{H}}(\mathcal{S}) & = \frac{1}{P}\operatorname{tr}(\widehat{\Sigma})   \\ 
    & = \frac{1}{P}\operatorname{tr}(\widehat{G}) - \frac{1}{P^2}\sum_{i,j=1}^P\widehat{G}_{ij} \\
    & = \frac{1}{P} \sum_{i=1}^P\frac{1}{N_i^2}\sum_{k=1}^{N_i}\sum_{l=1}^{N_i}\kappa(x^i_k, x^i_l)  \\
    & \quad - \frac{1}{P^2} \sum_{i=1}^{P}\sum_{j=1}^{P}\sum_{k=1}^{N_i} \frac{1}{N_iN_j}\kappa(x^i_k, x^j_l) \\
    & = \sum_{i=1}^{P}\sum_{j=1}^{P}\sum_{k=1}^{N_i} \sum_{l=1}^{N_j} A_{ijkl},
    \end{split}
\end{equation}
where 
\begin{equation}
    A_{i j k l}=\left\{\begin{array}{cc}     
\frac{P-1}{P^{2} N_{i}^{2}} \kappa \left(x_{k}^{i}, x_{l}^{j}\right) & i=j \\
\frac{-1}{P^{2} N_{i} N_{j}} \kappa \left(x_{k}^{i}, x_{l}^{j}\right) & i \neq j
\end{array}\right.
\end{equation}
\end{proof}

\newcommand{\code}[1]{\colorbox{gray!10}{\lstinline{#1}}}


\subsection{Reproducibility}    
\label{ap:rep}
\subsubsection{Dataset}    

\label{ap:dataset}     

To facilitate the reproduction of the results in our paper, we provide the processed Douban dataset in the supplementary material. Download and uncompress it to \code{./HRec/data/douban} folder. The dataset have been desensitized so that it contains no personally identifiable information or offensive content.  In this dataset, there are 1,096,240 interaction records between 303,825 items and 36,585 users. The interaction label in raw data is a rate from one-star to five-stars. For the convenience of comparison with some classic models, we convert the recommendation problem in the Douban dataset to a binary classification problem. The rates more than three-stars are treated as positive labels and others are treated as negative labels.

\subsubsection{Code}    

\label{ap:code}  

The code of our proposed model and baselines is available in Github\footnote{https://github.com/geekinglcq/HRec}. We provide the easy-to-use code so that anyone could run it with one simple line of code.

The code is written in \code{Python3.7}. We use the experience of RecBole \cite{recbole} for the implementation of part of the baselines.  

\textbf{Usage}~

The usage of our code is quite simple. Firstly, prepare a configuration file in JSON format. You can refer to the existing configuration files in the \code{configs} fold for examples.   

For instance, given a configuration file named \code{duration.json} in \code{configs}, just run the following command:  

\code{\$ python train_hete.py duration}    

For the homogeneous models, the usage is the same except for replacing \code{train_hete.py} with \code{train_homo.py}. Take the deepmf model as an example: 

\code{\$ python train_homo.py deepmf}

Once the program starts successfully, it creates a subfolder in \code{output}. The subfolder \code{\{Model name\}-\{Dataset name\}-\{time\}} contains the config file, the log file, and the best and the last checkpoints of the training model. 

\subsubsection{Setup}

In our offline experiments, the Douban dataset is divided into the train set (70\%), the validate set (20\%), and the test set (10\%). All of the models are trained with mini-batch Adam optimizer using early stopping criteria: if the performance of the model in the validation set does not improve within a patience threshold (five epochs here) then the training stops. Otherwise, the maximum number of training epochs is 100. The batch size is set to be 1024 and the initial learning rate is 0.001.  

After training, we select the checkpoint which performs best in the validate set and apply it in the test set to obtain the final result. As the evaluation metric, we use the Area Under Curve (AUC) to measure the probability that a model will assign a higher score to a randomly chosen positive item than a randomly chosen negative item. A higher AUC score indicates higher performance. We calculate the AUC for each individual kind of item along with the overall AUC of all items.

\begin{table*}[!bt] 
\begin{tabular}{ll|ll|ll}
\hline
\multicolumn{2}{c|}{AFM}                             & \multicolumn{2}{c|}{FISM}                      & \multicolumn{2}{c}{AutoInt}             \\ \hline
attention\_size          & 25                        & embedding\_size          & 64                  & embedding\_size   & 64                  \\
embedding\_size          & 10                        & split\_to                & 0                   & attention\_size   & 16                  \\
dropout\_prob            & 0.3                       & reg\_weights             & {[}0.01, 0.01{]}    & n\_layers         & 3                   \\
reg\_weight              & 2                         & alpha                    & 0                   & num\_heads        & 2                   \\ \cline{1-4}
\multicolumn{2}{c|}{DMF}                             & \multicolumn{2}{c|}{DeepFM}                    & dropout\_probs    & {[}0.2, 0.2, 0.2{]} \\ \cline{1-4}
user\_emb\_size          & 64                        & user\_emb\_size          & 64                  & mlp\_hidden\_size & {[}128, 128{]}      \\ \cline{5-6} 
item\_emb\_size          & 64                        & item\_emb\_size          & 64                  & \multicolumn{2}{c}{xDeepFM}             \\ \cline{5-6} 
user\_hidden\_size\_list & {[}64, 64{]}              & embedding\_size          & 64                  & user\_emb\_size   & 64                  \\
item\_hidden\_size\_list & {[}64, 64{]}              & mlp\_hidden\_size        & {[}128, 128, 128{]} & item\_emb\_size   & 64                  \\ \cline{1-2}
\multicolumn{2}{c|}{DDTCDR}                          & dropout\_prob            & 0.2                 & embedding\_size   & 10                  \\ \cline{1-4}
latent\_dim              & 32                        & \multicolumn{2}{c|}{CCCFNet}                   & mlp\_hidden\_size & {[}128, 128, 128{]} \\ \cline{3-4}
token\_emb\_size         & 32                        & user\_emb\_size          & 64                  & reg\_weight       & 0.0005              \\
layers                   & {[}64, 32{]}              & item\_emb\_size          & 64                  & dropout\_prob     & 0.2                 \\
alpha                    & 0.03                      & token\_emb\_size         & 32                  & direct            & FALSE               \\ \cline{1-2}
\multicolumn{2}{c|}{NAIS}                            & user\_hidden\_size\_list & {[}64, 64{]}        & cin\_layer\_size  & {[}100, 100, 100{]} \\ \cline{1-2} \cline{5-6} 
algorithm                & prod                      & item\_hidden\_size\_list & {[}64, 64{]}        & \multicolumn{2}{c}{DSSM}                \\ \cline{3-6} 
embedding\_size          & 64                        & \multicolumn{2}{c|}{Wide\&Deep}                & user\_emb\_size   & 64                  \\ \cline{3-4}
weight\_size             & 64                        & user\_emb\_size          & 64                  & item\_emb\_size   & 64                  \\
split\_to                & 0                         & item\_emb\_size          & 64                  & embedding\_size   & 10                  \\
reg\_weights             & {[}1e-07, 1e-07, 1e-05{]} & embedding\_size          & 64                  & mlp\_hidden\_size & {[}256, 256, 256{]} \\
alpha                    & 0                         & mlp\_hidden\_size        & {[}256, 64, 8{]}    & dropout\_prob     & 0.3                 \\
beta                     & 0.5                       & dropout\_prob            & 0.1                 & double\_tower     & TRUE                \\ \hline
\end{tabular}
\caption{Hyperparameters of baselines.}
\label{bbhp}
\end{table*}

 For a recommendation system, an important hyperparameter is the embedding size of users/items. We vary them in the set of \{32, 64, 128\} and observe that the optimal performance is obtained with 64. With regard to $\alpha$ and $\beta$ in Equation \ref{opt}, the selection of these two hyperparameters is related to the batch size due to our optimization method mentioned in Section \ref{obj}. Therefore, the guideline for tuning $\alpha$ and $\beta$ is to make the three loss terms in the same order of magnitude. When batch size is 1024, experimental results show that the setting of this model $\alpha=5\times 10^8, \beta=0.001$ leads  to the best performance. Overall, differences are minor indicating the insensibility of hyperparameters. We list the remaining hyperparameters of baselines in Table \ref{bbhp}.

\end{document}